\documentclass[onecolumn,aps]{revtex4}

\usepackage{graphicx}
\usepackage{dcolumn}
\usepackage{bm}

\begin{document}

\title
      {Superheavy element production, nucleus-nucleus potential and $\mu$-catalysis
      \footnote{To be published in proceedings of Tours 5 Symposium on
Nuclear Physics , August 26 – 29, 2003, see corresponding issue of
AIP Conference Proceedings, New York} }

\author{V.Yu. Denisov}

 \email{v.denisov@gsi.de; denisov@kinr.kiev.ua}
 \address{%
Gesellschaft f\"ur Schwerionenforschung (GSI), Planckstrasse 1,
D-64291 Darmstadt, Germany \\ Institute for Nuclear Research,
Prospect Nauki 47, 03028 Kiev, Ukraine}%

\begin{abstract}
The semi-microscopic potential between heavy nuclei is evaluated
for various colliding ions in the approach of frozen densities in
the framework of the extended Thomas-Fermi approximation with
$\hbar^2$ correction terms in the kinetic energy density
functional. The proton and neutron densities of each nucleus are
obtained in the Hartree-Fock-BCS approximation with SkM$^*$
parameter set of the Skyrme force. A simple expression for the
nuclear interaction potential between spherical nuclei is
presented. It is shown that muon bound with light projectile
induces the superheavy elements production in nucleus-nucleus
collisions.
\end{abstract}

\date{\today}

\maketitle

\section{Introduction}

Knowledge of the nucleus-nucleus interaction potential is a key
ingredient in the analysis of nuclear reactions. By using the
potential between nuclei we can estimate the cross sections of
different nuclear reactions \cite{bass}.

The nucleus-nucleus interaction potential related to the Coulomb
repulsion force and the nuclear attraction force has, as a rule,
the barrier and the capture potential well near a touching point.
The Coulomb part of the ion-ion potential is well-known. In
contrast, the nuclear part of the nucleus-nucleus potential is
less defined. There are many different approaches to the nuclear
part of the interaction potential [1-6]. Unfortunately, barriers
evaluated within different approaches for the same colliding
system differ considerably, especially when both nuclei are very
heavy or one nucleus is very heavy and another is light
\cite{dn,d-pot}. The uncertainty of the interaction potential
between heavy ions near the touching point gives rise to a variety
of proposed nuclear reaction mechanisms. So, there is a need to
reduce the uncertainty of the interaction potential around the
touching point, especially between heavy nuclei used for the
synthesis of superheavy elements (SHEs).

The production cross section of SHEs with $Z \geq 112$ is very low
and close to the limit of current experimental possibility [9-13].
Due to this it is of interest to find new types of reactions,
which can induce fusion of two heavy nucleus.

\begin{figure}
  \includegraphics[width=16.cm]{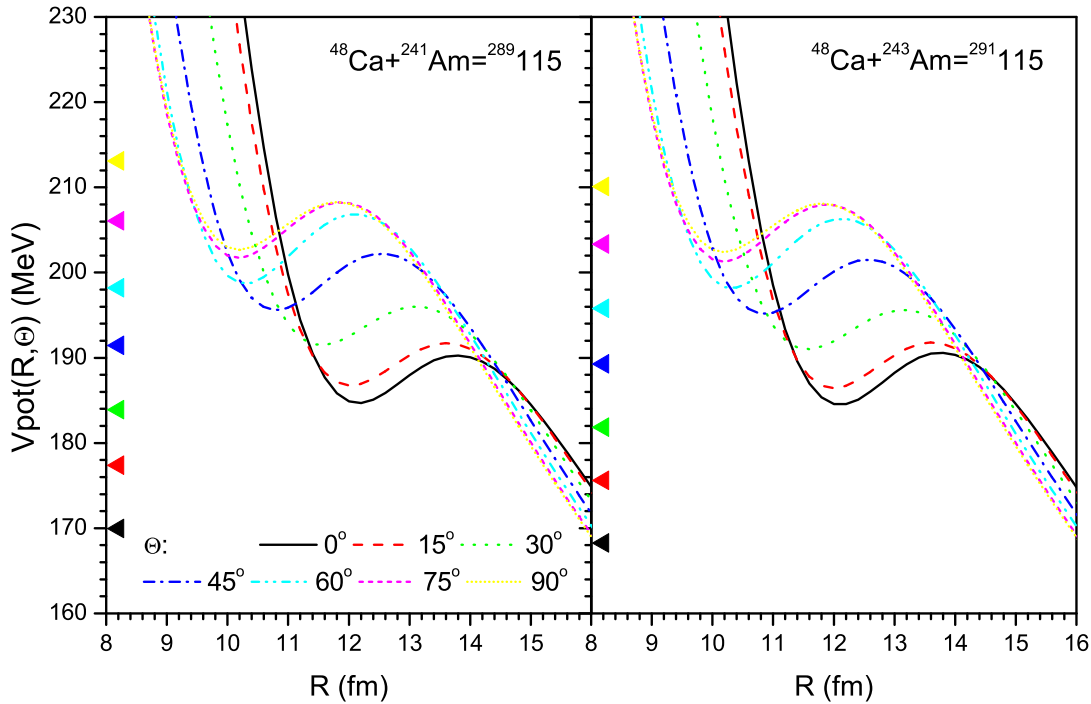}
  \caption{The SMPs for the collisions $^{48}$Ca on $^{241,243}$Am
evaluated with the SkM$^*$ Skyrme force. The SMPs are evaluated
for different angular orientations of the heavy deformed nuclei.
The ground-state Q-values are indicated by the lowest triangle at
the left vertical axis. The other 6 triangles mark are,
respectively, the thresholds for the emission of 1, 2, 3, 4, 5 and
6 neutrons.}
\end{figure}

In second section of the paper we briefly discuss our
semi-microscopic approach for the nucleus-nucleus interaction
potential and present some new numerical results. The simple
analytical expression for the nuclear potential between two heavy
spherical nuclei is presented in section 3. We show in section 4
that muon bound with light nucleus induce SHE formation during
nucleus-nucleus fusion reaction.

\section{Semi-microscopic potential and SHE production}

We evaluate the nuclear part of interaction potential between
heavy nuclei in the semi-microscopic frozen density approximation
due to a short reaction time \cite{dn}. The frozen (or sudden)
approximation is good for evaluation of the nucleus-nucleus
potential near the touching point at collision energies above the
barrier height. The shape of each ion cannot appreciably change
and the energy of relative motion cannot be strongly transferred
to another degrees of freedom during the short reaction time.

\begin{figure}
  \includegraphics[width=16.cm]{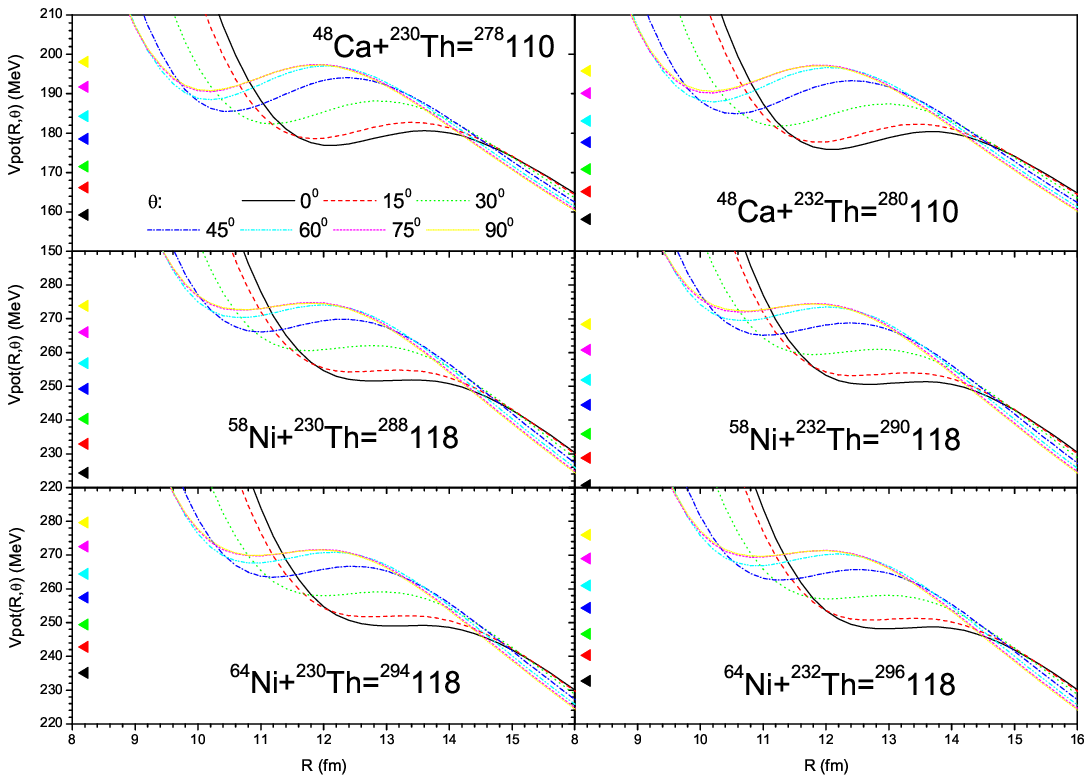}
  \caption{The SMPs for the collisions $^{48}$Ca on $^{230,232}$Th
and $^{58,64}$Ni on $^{230,232}$Th evaluated with the SkM$^*$
Skyrme force. The notations are the same as in Fig. 1.}
\end{figure}

The interaction energy between ions is obtained with the help of a
local energy density functional. The extended Thomas-Fermi (ETF)
approximation with $\hbar^2$ correction terms is used for the
evaluation of the kinetic energy density functional \cite{brack}.
The Skyrme and Coulomb energy density functionals are employed for
the calculation of the potential energy. These energy density
functionals depend on the proton and neutron densities. These
densities in each nuclei are obtained in the microscopic
Hartree-Fock-BCS approximation with the Skyrme force. Our
approximation is semi-microscopic because we use the microscopic
density distributions and the ETF approximation for the
calculation of the interaction energy of nuclei. Note that the
binding energies of nuclei evaluated in the ETF approximation with
the help of microscopic density distributions well agree with
those obtained in the fully microscopic Hartree-Fock-BCS model
\cite{dn,d-pot}. Therefore, our semi-microscopic method for
evaluation of the interaction potential between various nuclei is
quite accurate. The details of our method and  results for various
even-even projectiles and targets are discussed in [7-8,15-16].
Due to this we present here only some new results obtained in our
approximation and discussions.

At the beginning we consider the semi-microscopic potentials
(SMPs) for hot fusion reactions which was used for the formation
of element with 115 protons in Dubna very recently \cite{114-116}.
In Fig. 1 we presented SMPs for reactions
$^{48}$Ca+$^{241,243}$Am=$^{289,291}115$ evaluated for SkM$^*$
Skyrme force \cite{skm}. We see that the potential shape (radial
positions of both the barriers and the capture wells for various
orientations of deformed $^{241,243}$Am and the depths of capture
wells) for these reactions are very similar to the ones for
reactions between $^{48}$Ca and other closest even-even targets
$^{238}$U, $^{236,238,240,242,244}$Pu, $^{248}$Cm and $^{252}$Cf
\cite{dn,d-slovakia,d-magdeburg}. However the positions of both
the barriers and the capture wells relatively the ground state of
compound nucleus for various orientations $^{241,243}$Am are
slightly lower then the ones evaluated for nearest even-even
projectiles $^{240,242,244}$Pu, see Fig. 1 in \cite{d-magdeburg}.
This is favorable for the SHE formation in reaction with
$^{241,243}$Am.

A choice of light isotope is very important for the SHE production
\cite{dn,d-slovakia,d-magdeburg}. A SHE formation reactions
induced by $^{48}$Ca projectile are limited by availability of
heavy transuranic elements in the nature. Due to this it is
practically impossible to synthesize element with $Z \geq 120$ by
using $^{48}$Ca beam. By using Fig. 2 and Figs. 4-7 in \cite{dn}
and analyzing the chapter of nuclides we may conclude that
$^{64}$Ni is the nearest nuclide heavier then $^{48}$Ca, which we
may recommend to use for the SHE production, because of
\begin{itemize}
\item[{\it (1)}] $^{64}$Ni is easily available as a projectile,

\item[{\it (2)}] $^{64}$Ni is sufficiently neutron reach,

\item[{\it (3)}] $^{64}$Ni produces relatively low-excited
compound nuclei during fusion with easily available set of target
nuclei,

\item[{\it (4)}] the number of protons in $^{64}$Ni is magic as
that in $^{48}$Ca.

\end{itemize}
Therefore, reactions with $^{64}$Ni projectile
($^{64}$Ni+X$\rightarrow$SHE) should play similar role as
reactions with $^{48}$Ca ($^{48}$Ca+X$\rightarrow$SHE), which are
successfully used now in Dubna \cite{114-116}. However the capture
well for reactions with $^{64}$Ni is more shallow then the one for
reactions with $^{48}$Ca, see Fig. 2. This is may reduce the
capture probability for reactions with $^{64}$Ni projectile.

By analyzing SMPs for reactions presented in Figs. 1,2 and for
other reactions considered in \cite{dn,d-slovakia,d-magdeburg} we
observe the common features of SMPs for various reactions and make
conclusions for reactions leading to the SHEs:
\begin{itemize}

\item[{\it (1)}] The depth of the pockets is important for the
fusion probability. We observe correlation between pocket depth
and experimentally observed reduction of SHE formation with
increasing size of the projectile.

\item[{\it (2)}] The pocket depth should be as large as possible
for the SHE production, because the deeper pocket has larger the
capture window and, therefore, better chance for fusion. Due to
this hot fusion reactions has better chance for capture then cold
fusion reactions leading to the same SHE.

\item[{\it (3)}] For the subsequent formation of a compound
nucleus it is best to have a most compact capture configuration.
Due to this both the cold fusion systems are more preferable then
more symmetric systems leading to the same compound nucleus and
the side orientation of deformed nucleus is more preferable for
the SHE formation then the tip orientation of deformed nucleus in
the case of the hot fusion reactions.

\item[{\it (4)}]  The observed fusion windows lie systematically
about 5 to 10 MeV below our barriers.

\item[{\it (5)}] The isotopic composition of heavy transuranic
nuclei only weakly affects both the shape of the capture well and
barrier heights for different orientations relatively the
ground-state fusion reaction $Q$-value.

\item[{\it (6)}] The capture properties of SMP and
compound-nucleus excitation energy in the hot fusion reactions
strongly depend on the isotopic composition of light nuclide.

\item[{\it (7)}]  The difference between the capture barrier
position and the ground-state fusion reaction $Q$-value decreases
with increasing as charge as number of neutrons of the projectile
for reactions with the same target.

\item[{\it (8)}]  Pocket depth is vanished with increasing size of
the projectile. For example in the case of $^{208}$Pb target,
there are no pocket for $^{96}$Zr+$^{208}$Pb and more heavy
projectiles.

\item[{\it (9)}]  The potential pockets for system leading to SHE
are much shallower than the ones for more lighter colliding
systems.

\item[{\it (10)}]  The interaction potentials obtained by using
different standard expressions [1-6] are spread over large
interval for heavier systems.

\end{itemize}

\section{Analytical expression for potential}

The numerical evaluation of SMP is very accurate for determination
of interaction potential between nuclei around the touching point.
Unfortunately, it is not so convenient for any practical
application because one needs to evaluate numerically the
microscopic Hartree-Fock-BCS nucleonic densities of interacting
nuclei, derivatives of these densities and integrals
\cite{dn,d-pot}. It is better to find analytical expression for
the potential. To solve this task we choose 119 spherical or near
spherical nuclei along the $\beta$-stability line from $^{16}$O to
$^{212}$Po and perform numerical calculations of the interaction
potentials between all possible nucleus-nucleus combinations in
the semi-microscopic approximation. We evaluate potential for any
nucleus-nucleus combinations at 15 distances between ions around
the touching point. By using database for 7140 nucleus-nucleus
potentials at 15 points each, we find a simple analytical
expression for the potential between spherical nuclei in the form
\cite{d-pot}
\begin{eqnarray}
V(R)= - 1.989843 \; C \; f(R-R_{12}-2.65) \\ \times
\left[1+0.003525139 (A_1/A_2+A_2/A_1)^{3/2} - 0.4113263
(I_1+I_2)\right] , \nonumber
\end{eqnarray}
where $R$ is the distance between mass centers of colliding
nuclei, $C=R_1 R_2 / R_{12}$, $R_i$ is the effective nuclear
radius, $R_{12}=R_1+R_2$,
\begin{eqnarray}
f(s)=\left\{1 - s^2\left[0.05410106 \; C
\exp\left(-\frac{s}{1.760580}\right) - 0.5395420 \; (I_1+I_2)
\right. \right. \;\;\;\;\; \\ \left. \left. \times
\exp\left(-\frac{s}{2.424408}\right)\right]\right\} \times
\exp\left(\frac{-s}{0.7881663}\right), \;\; {\rm for}
\;\; s \geq 0, \;\;  \nonumber \\
f(s)= 1-\frac{s}{0.7881663}+1.229218 s^2-0.2234277s^3-0.1038769s^4
\;\;\;\;\;\;\; \\
-C(0.1844935s^2+0.07570101s^3)  \nonumber
\\+(I_1+I_2)(0.04470645s^2+0.03346870 s^3), \;\;{\rm for}
\;\; -5.65 \leq s \leq 0, \; \nonumber
\end{eqnarray}
$A_i$ is the number of nucleon in nucleus $i$ ($i=1,2$),
$I_i=(N_i-Z_i)/A_i$, $Z_i$ and $N_i$ are numbers of protons and
neutrons in nucleus $i$. The effective nuclear radius is given by
\begin{equation}
R_i=R_{ip} (1-3.413817 /R_{1p}^2) + 1.284589
(I_i-0.4A_i/(A_i+200)),
\end{equation}
where the proton radius is determined as in \cite{pomorska}
$R_{ip}=1.24 A_i^{1/3} (1+1.646/A_i-0.191 I_i)$.
The last term in (4) takes into account deviation of the nuclear
radius from the proton radius when the neutron number in nucleus
deviates from the $\beta$-stability value for given $A$. The line
of $\beta$-stability is described by Green's approximation $I =
(N-Z)/A = 0.4 A/(A+200)$ \cite{green}. Note that the potentials
obtained by means of the analytical expression well agree with
semi-microscopic one \cite{d-pot}.

\begin{figure}
  \includegraphics[width=15.cm]{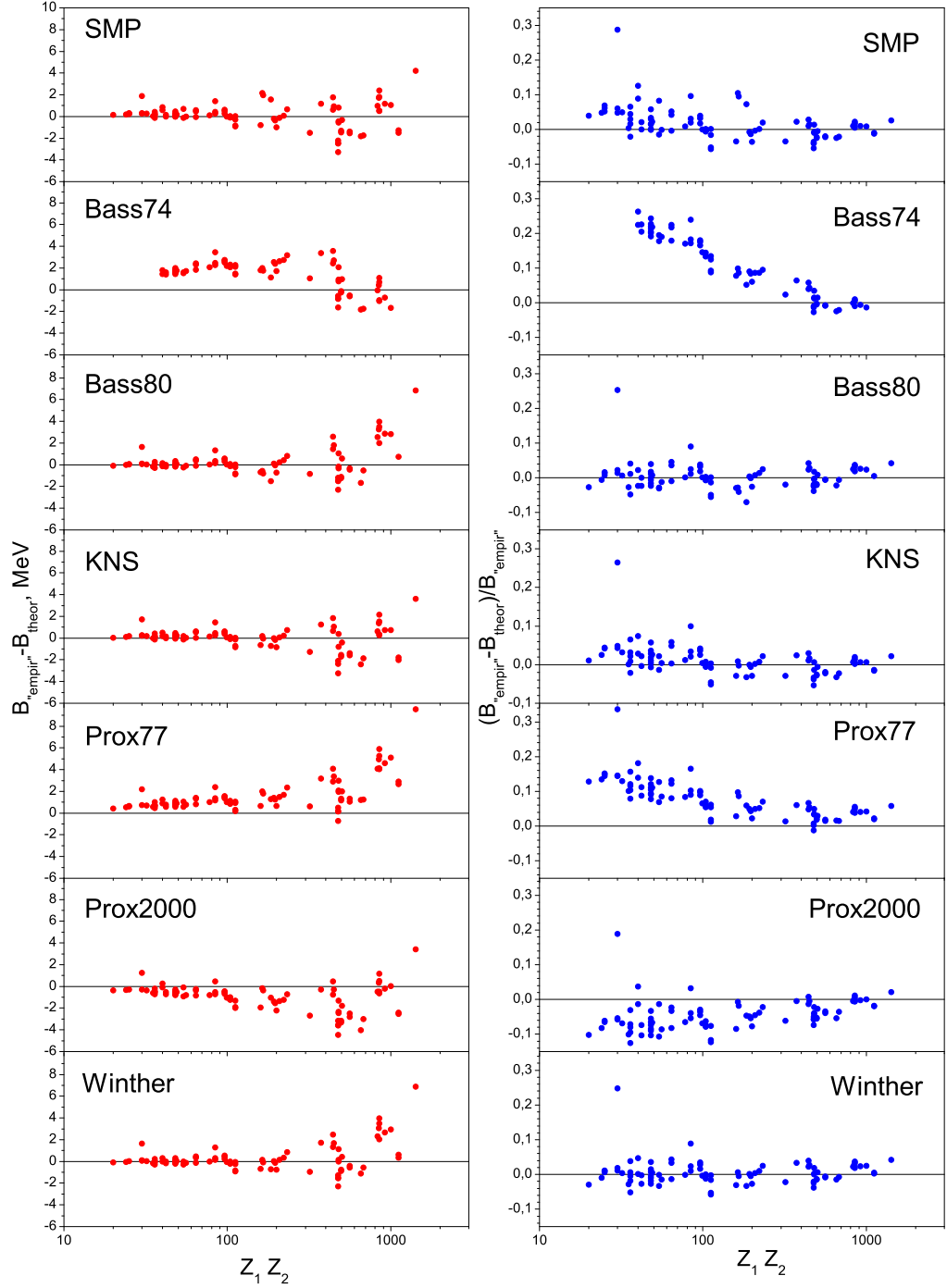}
  \caption{The absolute (left panels) and relative (right panels)
  differences between the "empirical" fusion barrier and
the barrier evaluated by using various analytical expressions (SMP
- \cite{d-pot}, Bass74 - \cite{bass74}, Bass80 - \cite{bass},
Prox77 - \cite{prox77} , Prox2000 - \cite{prox2000}, KNS -
\cite{kns} and Winther - \cite{winther}).}
\end{figure}

\begin{figure}
  \includegraphics[width=15.cm]{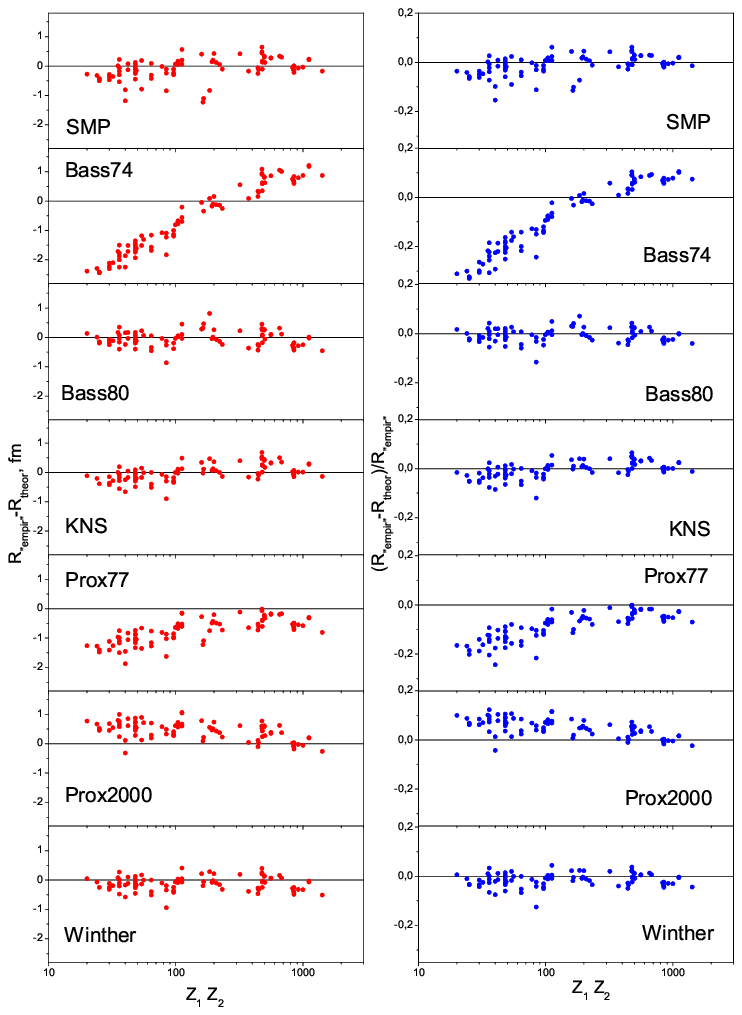}
  \caption{The absolute (left panels) and relative (right panels)
  differences between the "empirical" radius of barrier and
the barrier radius evaluated by using various analytical
expressions (SMP - \cite{d-pot}, Bass74 - \cite{bass74}, Bass80 -
\cite{bass}, Prox77 - \cite{prox77} , Prox2000 - \cite{prox2000},
KNS - \cite{kns} and Winther - \cite{winther}).}
\end{figure}

The "empirical" fusion barrier $B_{\rm empir}$ and "empirical"
barrier radius $R_{\rm empir}$ between heavy nuclei are extracted
by means of a special analysis of the experimental data for
subbarrier fusion reactions in Ref. \cite{bar}. The absolute and
relative differences between "empirical" fusion barrier and
barrier evaluated by using various analytical expressions [1-6,8]
$B_{\rm theor}$ are presented in Fig. 3. In Fig. 4 we present the
absolute and relative differences between "empirical" barrier
radius and radii of barrier evaluated by using various analytical
expressions [1-6,8] $R_{\rm theor}$. The barriers $B_{\rm theor}$
and radii $R_{\rm theor}$ evaluated by using Eqs. (1)-(4) well
agree with "empirical" the ones respectively, see Figs. 3-4. The
distributions of deviations $B_{\rm empir}-B_{\rm theor}$,
$(B_{\rm empir}-B_{\rm theor})/B_{\rm empir}$, $R_{\rm
empir}-R_{\rm theor}$ and $(R_{\rm empir}-R_{\rm theor})/R_{\rm
empir}$ are almost symmetric with respect to the lines $B_{\rm
empir}-B_{\rm theor} =0$ or $R_{\rm empir}-R_{\rm theor}=0$
correspondingly for the case of using analytical expression for
SMP, see Figs. 3-4. This also suggests the reliability of $A$- and
$Z$-dependencies of our expression for the nucleus-nucleus
potential. In contrast to this similar distributions for the
proximity-1977 \cite{prox77}, proximity-2000 \cite{prox2000} and
Bass-1974 \cite{bass74} potentials have no symmetry with respect
to the lines $B_{\rm empir}-B_{\rm theor} =0$ and $R_{\rm
empir}-R_{\rm theor}=0$, see corresponding panels in Figs. 3-4.
Note that it is impossible to evaluate both the depth and the
width of capture well by using Bass-1974 \cite{bass74}, Bass-1980
\cite{bass}, Krappe-Nix-Sierk \cite{kns} and Winther
\cite{winther} potentials because the shape of these potentials
are unrealistic or unknown at distances smaller then the touching
point distance of two nuclei.

\section{Catalysis of the SHE synthesis by muon}

It is easy to understand qualitatively a influence of muon $\mu^-$
on the SHE fusion process, if we recollect that the wave function
of $1s$ state of $\mu^-$ in a very heavy nucleus is located inside
the nucleus \cite{kim}. Therefore, negatively-charged muon inside
heavy nucleus should effectively reduce the Coulomb repulsion
between protons. Due to this the forces, inducing fission of
compound nucleus and preventing fusion of two nuclei should
decrease. Consequently, the SHE formation probability should rise
due to $\mu^-$.

Let's we consider in detail the catalysis of the SHE production by
$\mu^-$. The process of SHE formation is subdivided into three
steps [22-27]: \begin{itemize}

\item[{\it (1)}] The capture of two nuclei in an entrance-channel
potential well and formation of a common nuclear system of two
touching nuclei.

\item[{\it (2)}] The formation of a spherical or nearly spherical
compound nucleus during shape evolution from the common nuclear
system of two touching nuclei to a compound nucleus.

\item[{\it (3)}] The surviving of the excited compound nucleus due
to evaporation of neutrons and $\gamma$-ray emission in
competition with fission.

\end{itemize}

The capture process depends on both the barrier thickness and
pocket shape of entrance-channel potential between nuclei
\cite{dn}. The shape evolution step is determined by a
potential-energy landscape between the touching configuration of
two colliding nuclei and the compound-nucleus shape [22-27]. Decay
properties of the compound nucleus drastically depend on the
fission barrier height [22-27]. Therefore, enhancement of the SHE
production in fusion reaction may be achieved by processes which
\begin{itemize}
\item[{\it (1)}] make capture pocket dipper and barrier of
entrance-channel potential thinner, \item[{\it (2)}] increase the
slope or reduce both barrier height and thickness of the
potential-energy landscape between touching configuration of two
colliding nuclei and compound nucleus, \item[{\it (3)}] increase
the fission barrier height.
\end{itemize}
Below we show that these three conditions can be met in a reaction
between a light nucleus with captured $\mu^-$-meson $L_\mu$ and a
heavy nucleus $T$.

The potential energy of $L_\mu$+$T$ system before touching can be
approximated as
\begin{equation}
E_{L_\mu T}(R) = B_L + B_T + B_{L\mu} + V_{LT}(R) + V_{T\mu}(R),
\end{equation}
where $B_L$ and $B_T$ are the binding energies of light $L$ and
heavy $T$ nuclei, respectively, $B_{L\mu}$ is the binding energy
of muon in the light nucleus $L$, $V_{LT}(R)$ is the interaction
potential between the light and heavy nuclei related to Coulomb
and nuclear forces at distance $R$ between their mass centers
\cite{dn,d-pot}, and $V_{T\mu}(R)=-e^2 Z_T/R$ is the Coulomb
interaction between $Z_T$ protons in the heavy nucleus and the
muon. The potential energy of the compound nucleus with bound
$\mu^-$ is connected with the binding energy of the compound
nucleus $B_{CN}$ and with that of muon in the compound nucleus
$B_{CN\mu}$, i.e., $E_{CN} = B_{CN} + B_{CN\mu}$.

The potential-energy landscape evaluated relatively to the ground
state of compound nucleus with bound $\mu^-$, which formed during
$L_\mu$+$T$ fusion reaction, is related to
\begin{equation}
\delta(R)=E_{L_\mu T}(R)-E_{CN} = \delta_N(R) + \delta_{N\mu}(R).
\end{equation}
Here we split $\delta(R)$ into contributions of pure nuclear
$\delta_N(R)$ and muon-nuclear $\delta_{N\mu}(R)$ subsystems,
where
\begin{eqnarray}
\delta_N(R) = B_L + B_T -  B_{CN} + V_{LT}(R), \;\;\;\;\;\;\;\;
\delta_{N\mu}(R) = B_{L\mu} - B_{CN\mu} + V_{T\mu}(R).
\end{eqnarray}
We see that the Coulomb interaction between muon and protons
modifies the potential-energy landscape of  fusing system. (Note
that realistic landscape of potential-energy surface of fusing
system depends on a great number of various collective
coordinates. However, we take into account only the most important
collective coordinate, which describes the distance between mass
centers of separated nuclei or elongation of fusing system upon
the capture step.)

Note that $ \delta_{N\mu}(R) = B_{L\mu} - B_{CN\mu} - Z_T e^2/R$
is decreased with reduction of $R$. At distance $R_{CN}$, which
corresponds to the distance between left and right mass centers of
compound nucleus, $\delta_N(R_{CN})=\delta_{N\mu}(R_{CN})=0$
because we evaluate $\delta(R)$ relatively to the ground state of
compound nucleus with bound $\mu^-$. Note that $B_{L\mu} -
B_{CN\mu} > 0$, therefore $\delta_{N\mu}(R)$ is mainly reduced
with $R$. Consequently if $\delta_{N\mu}(R)$ continuously
decreases with reducing of $R$, then muon induces the SHE
formation due to three effects:
\begin{itemize}

\item[{\it ( 1)}] A more dipper capture pocket is formed as a
result of such $R$ dependence of $\delta_{N\mu}(R)$. Therefore,
the capture state formation probability increases.

\item[{\it ( 2)}] The potential-energy landscape of the
muon-nuclear system becomes more favorable for shape evolution
from captured states of two touching nuclei to the compound
nucleus.

\item[{\it ( 3)}] The muon-nuclear system exhibits a larger
fission barrier height as compared to pure nuclear system, see
also \cite{mu_fis} and papers cited therein. Note that variation
of the fission barrier height on 0.3 MeV lead to change of SHE
production cross section on approximately 30\% \cite{dh}. Due to
muon the fission barrier of heavy nucleus is increased near to 1
MeV \cite{mu_fis}. Consequently, the fission or quasi-fission
probability of muon-nuclear system get reducing approximately on
one order as compared to the pure nuclear system. Therefore only
due to this effect the SHE production cross section may rise to
one order.

\end{itemize}

$\mu^-$ is a convenient particle for inducing compound-nucleus
formation in reactions $L_\mu + T \rightarrow SHE + xn + e^- +
\overline{\nu}_e + \nu_\mu$, because its lifetime ($\approx
2.2\times 10^{-6}$s \cite{kim}) is sufficient for making $1s$
bound state with a light projectile nucleus just before the
collision with a target and induce fusion reaction. The process of
SHE formation during nucleus-nucleus collision is fast relatively
typical $\mu^-$ dynamic time. Therefore there is high probability
of population of $1s$ bound state of $\mu^-$ in SHE during nuclear
reaction time. The compound nucleus relatively rarely excited
during the decay $\mu^-$ ($\mu^- \rightarrow e^- +
\overline{\nu}_e + \nu_\mu$ \cite{kim}). It is possible to make
beam of muonic projectile $L_\mu$ by merging beams of strongly
ionized projectile nucleus $L$ and of $\mu^-$ at the same
velocities before the target.

Note that muon catalysis of thermonuclear reactions is also
related to effective reduction of the Coulomb repulsion between
protons and is well studied both theoretically and experimentally
(see \cite{kim} and papers cited therein). Muon catalysis of
thermonuclear reactions between two hydrogen isotopes is mainly
related to reduction of both fusion barrier heights and thickness.
In contrast to this muon catalysis of SHE production is connected
with more complex processes as reduction of fusion barrier
thickness, modification of capture pocket, variation of
potential-energy landscape between capture and compound-nucleus
shapes and rising of the fission barrier height.

The author would like to thank W. N\"orenberg and Yu. Ts.
Oganessian for useful discussions and communications. Author
gratefully acknowledges support from GSI.

\end{document}